\documentclass[epj]{webofc-mod}
\usepackage[varg]{txfonts}   
\usepackage[hyperindex,breaklinks]{hyperref}
\usepackage{url} 
\usepackage{subfigure}
\def\to{\ensuremath{\rightarrow}}

\DeclareGraphicsExtensions{.pdf}
\woctitle{New Frontiers in Physics 2014}
\begin{document}
\hfill IFIC/14-66
\title{The physics case for the MoEDAL experiment at LHC}
%

\author{Vasiliki A.\ Mitsou\inst{1}\fnsep\thanks{\email{vasiliki.mitsou@ific.uv.es}}\\ On behalf of the MoEDAL Collaboration} 

\institute{Instituto de F\'isica Corpuscular (IFIC), CSIC -- Universitat de Val\`encia, \\ 
Parc Cient\'ific de la U.V., C/ Catedr\'atico Jos\'e Beltr\'an 2, \\
E-46980 Paterna (Valencia), Spain}

\abstract{%
The MoEDAL experiment (Monopole and Exotics Detector at the LHC) is designed to directly search for magnetic monopoles and other highly-ionising stable or metastable particles arising in theoretical scenarios beyond the Standard Model. Its physics goals are accomplished by the deployment of plastic nuclear track detectors combined with trapping volumes for capturing charged highly-ionising particles and TimePix pixel devices for monitoring. This paper is an overview of the MoEDAL physics reach, which is largely complementary to the programs of the large multi-purpose LHC detectors ATLAS and CMS. }
\maketitle
%
\section{Introduction}\label{sc:intro}

MoEDAL (Monopole and Exotics Detector at the LHC)~\cite{moedal-web,moedal-tdr,jim}, the $7^{\rm th}$ experiment at the Large Hadron Collider (LHC)~\cite{LHC}, was approved by the CERN Research Board in 2010. It is designed to search for manifestations of new physics through highly-ionising particles in a manner complementary to ATLAS and CMS~\cite{DeRoeck:2011aa}. The most important motivation for the MoEDAL experiment is to pursue the quest for magnetic monopoles and dyons at LHC energies. Nonetheless the experiment is also designed to search for any massive, stable or long-lived, slow-moving particles~\cite{Fairbairn07} with single or multiple electric charges arising in many scenarios of physics beyond the Standard Model (SM). A selection of the physics goals and their relevance to the MoEDAL experiment are described here. For an extended and detailed account of the MoEDAL discovery potential, the reader is referred to the recently published \emph{MoEDAL Physics Review}~\cite{Acharya:2014nyr}.

The structure of this paper is as follows. Section~\ref{sc:detector} provides a brief description of the MoEDAL detector. The physics reach of MoEDAL as far as magnetic monopoles and monopolia is discussed in Section~\ref{sc:mm}, whilst Section~\ref{sc:susy} is dedicated to supersymmetric models predicting massive (meta)stable states. Scenarios with doubly-charged Higgs bosons and their observability in MoEDAL are highlighted in Section~\ref{sc:lrsm}. Highly-ionising exotic structures in models with extra spatial dimensions, namely microscopic black holes and D-matter, relevant to MoEDAL are briefly mentioned in Section~\ref{sc:ed}. The paper concludes with a summary and an outlook in Section~\ref{sc:summary}.

\section{The MoEDAL detector}\label{sc:detector}

\begin{figure}[htb]
\begin{center}
\includegraphics[width=0.62\textwidth]{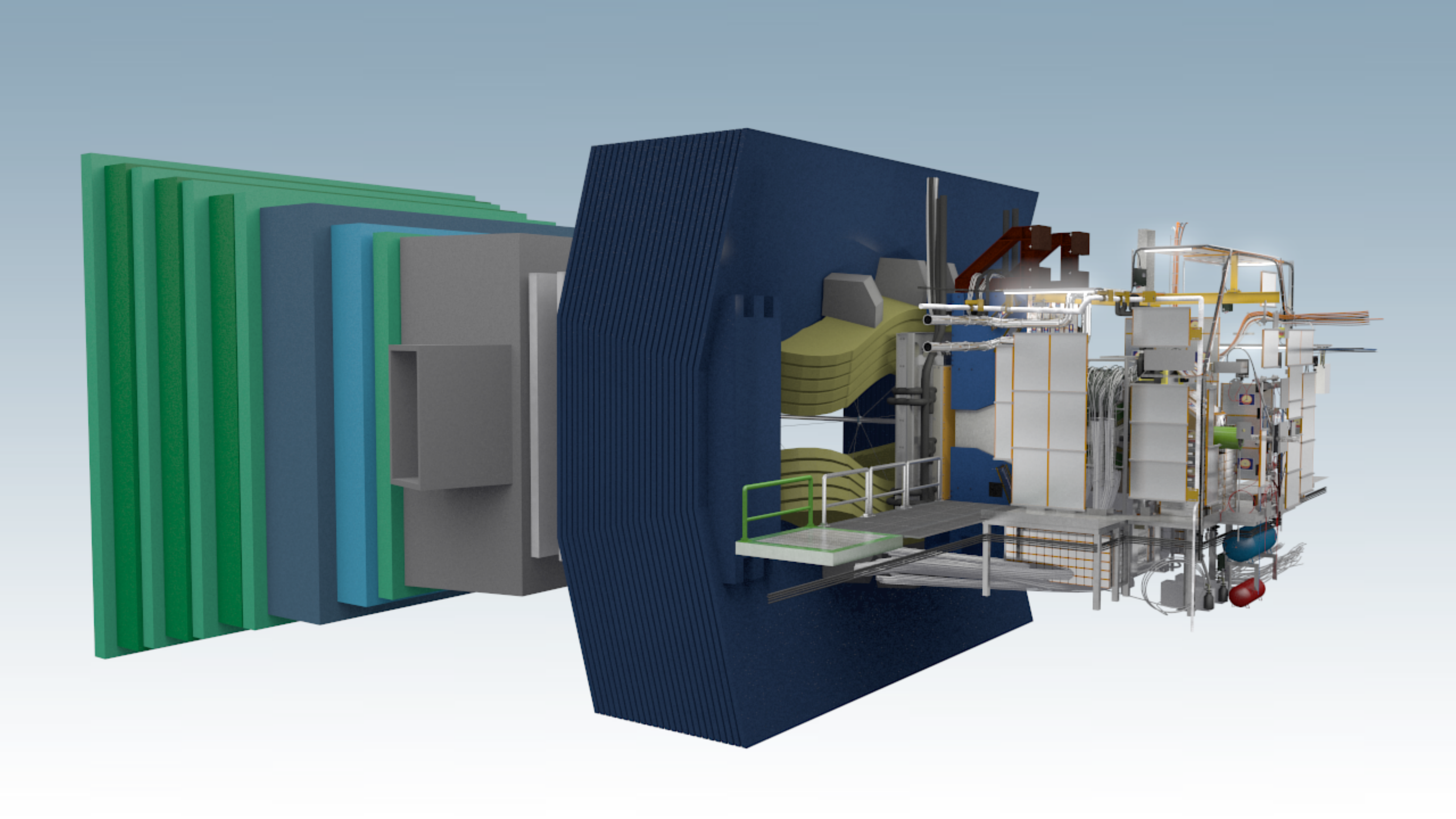}
\caption{ A three-dimensional schematic view of the MoEDAL detector (on the right) around the LHCb VELO region at Point~8 of the LHC.}
\label{Fig:moedal-lhcb}
\end{center}
\end{figure}

The MoEDAL detector~\cite{moedal-tdr} is deployed around the intersection region at Point~8 of the LHC in the LHCb experiment Vertex Locator (VELO)~\cite{LHCb-detector} cavern. A three-dimensional depiction of the MoEDAL experiment is presented in Fig.~\ref{Fig:moedal-lhcb}. It is a unique and largely passive LHC detector comprised of four sub-detector systems. 

The main sub-detector system is made of a large array of CR39\textregistered,  Makrofol\textregistered\ and Lexan\textregistered\ nuclear track detector (NTD) stacks surrounding the intersection area. The passage of a highly-ionising particle through the plastic detector is marked by an invisible damage zone along the trajectory. The damage zone is revealed as a cone-shaped etch-pit when the plastic detector is etched using a hot sodium hydroxide solution. Then the sheets of plastics are scanned looking for aligned etch pits in multiple sheets. The MoEDAL NTDs have a threshold of $Z/\beta\sim5$, where $Z$ is the charge and $\beta=v/c$ the velocity of the incident particle. In proton-proton collision running, the only source of known particles that are highly ionising enough to leave a track in MoEDAL NTDs are spallation products with range that is typically much less than the thickness of one sheet of the NTD stack. In that case the ionising signature will be that of a very low-energy electrically-charged \emph{stopped} particle. This signature is distinct to that of a  \emph{penetrating} electrically or magnetically charged particle that will usually traverse every sheet in a MoEDAL NTD stack, accurately demarcating a track that points back to the collision point. During the heavy-ion running one might expect a background from high ionising fragments, which however are produced in the far forward direction and do not fall into the acceptance of the MoEDAL detector.

A unique feature of the MoEDAL detector is the use of paramagnetic magnetic monopole trappers (MMTs) to capture electrically- and magnetically-charged highly-ionising particles. The aluminium absorbers of MMTs will be subject to an analysis looking for magnetically-charged particles at a remote magnetometer facility~\cite{Joergensen:2012gy,DeRoeck:2012wua}. The search for the decays of long-lived electrically charged particles that are stopped in the trapping detectors will subsequently be carried out at a remote underground facility.

The only non-passive MoEDAL sub-detector system comprises an array of around ten TimePix pixel devices that form a real-time radiation monitoring system dedicated to the monitoring of highly-ionising background sources in the MoEDAL cavern.

\section{Magnetic monopoles}\label{sc:mm}

The MoEDAL detector is designed to fully exploit the energy-loss mechanisms of magnetically charged particles~\cite{Dirac1931kp,Diracs_idea,tHooft-Polyakov,mm,Cho1996qd}  in order to optimise its potential to discover these messengers of new physics. There are various theoretical scenarios in which magnetic charge would be produced  at the LHC~\cite{Acharya:2014nyr}: (light) 't Hooft-Polyakov monopoles~\cite{tHooft-Polyakov,Vento2013jua}, electroweak monopoles~\cite{Cho1996qd} and monopolium~\cite{Diracs_idea,khlopov,Monopolium,Monopolium1}. Magnetic monopoles that carry a non-zero magnetic charge and dyons possessing both magnetic and electric charge are among the most fascinating  hypothetical particles. Even though there is no generally acknowledged empirical  evidence for their existence, there are strong theoretical reasons to believe that they do exist, and they are predicted by many theories including grand unified theories and superstring theory~\cite{Rajantie:2012xh}. 
 

The theoretical motivation behind the introduction of magnetic monopoles is the symmetrisation of the Maxwell's equations and the explanation of the charge quantisation~\cite{Dirac:1931kp}. Dirac showed that the mere existence of a monopole in the universe could offer an explanation of the discrete nature of the electric charge, leading to the Dirac Quantisation Condition (DQC),

\begin{equation} e  g = \frac{N}{2} , \quad  N = 1, 2, ... , 
\label{eq:dqc}\end{equation}

\noindent where $e$ is the electron charge and $g$ the monopole magnetic charge. In Dirac's formulation, magnetic monopoles are assumed to exist as point-like particles and quantum mechanical consistency conditions lead to Eq.~(\ref{eq:dqc}), establishing the value of their magnetic charge. Although monopoles symmetrise Maxwell's equations in form, there is a numerical asymmetry arising from the DQC, namely that the basic magnetic charge is much larger than the smallest electric charge. A magnetic monopole with a single Dirac charge ($g_{\rm D}$) has an equivalent electric charge of  $\beta(137e/2)$. Thus  for a relativistic monopole the energy loss is around $4,700$ times ($68.5^2$) that of a minimum-ionising electrically-charged particle. The monopole mass remains a free parameter of the theory.


No magnetically-charged particles have been observed so far. A possible explanation for this lack of experimental confirmation is Dirac's proposal~\cite{Dirac:1931kp,Diracs_idea,khlopov} that monopoles are not seen freely because they form a bound state called \emph{monopolium}~\cite{Monopolium,Monopolium1,Epele0} being confined by strong magnetic forces. Monopolium is a neutral state, hence it is difficult to detect directly at a collider detector, although its decay into two photons would give a rather clear signal for the ATLAS and CMS detectors~\cite{Epele1}, which however would not be visible in the MoEDAL detector. Nevertheless the monopolium might break up in the medium of MoEDAL into highly-ionising dyons, which subsequently can be detected in MoEDAL~\cite{Acharya:2014nyr}. Moreover its decay via photon emission would produce a peculiar trajectory in the medium, should the decaying states are also magnetic multipoles~\cite{Acharya:2014nyr}.

\section{Electrically-charged long-lived particles in supersymmetry}\label{sc:susy}

Supersymmetry (SUSY)~\cite{susy} is an extension of the Standard Model which assigns to each SM field a superpartner field with a spin differing by a half unit. SUSY provides elegant solutions to several open issues in the SM, such as the hierarchy problem, the identity of dark matter, and the grand unification. SUSY scenarios propose a number of massive slowly moving ($\beta \lesssim 5$)  electrically charged particles. If they  are sufficiently long-lived to  travel a distance of at least ${\cal O}(1~{\rm m})$  before decaying and their $Z/\beta\gtrsim 0.5$,  then they will be detected in the MoEDAL NTDs. No highly-charged particles are expected in such a theory, but there are several scenarios in which supersymmetry may yield massive, long-lived particles that could have electric charges $\pm 1$, potentially detectable in MoEDAL if they are produced with low velocities.

The lightest supersymmetric particle (LSP) is stable in models where $R$~parity is conserved. The LSP should have no strong or electromagnetic interactions, for otherwise it would bind to conventional matter and be detectable in anomalous heavy nuclei~\cite{EHNOS}. Possible weakly-interacting neutral candidates in the Minimal Supersymmetric Standard Model (MSSM) include the sneutrino, which has been excluded by LEP and direct searches, the lightest neutralino $\tilde{\chi}_1^0$ (a mixture of spartners of the $Z, H$ and $\gamma$) and the gravitino.

\subsection{Supersymmetric scenarios with $R$-parity violation}

Several scenarios featuring metastable charged sparticles might be detectable in MoEDAL. One such scenario is that $R$~parity {\it may not be exact}~\cite{RV}, since there is no exact local symmetry associated with either $L$ or $B$, and hence no fundamental reason why they should be conserved. One could consider various ways in which $L$ and/or $B$ could be violated in such a way that $R$ is violated, as represented by the following superpotential terms~\cite{RV}:
\begin{equation}
W_{RV} \; = \; \lambda^{\prime \prime}_{ijk} {\bar U}_i {\bar D}_j {\bar D}_k
+  \lambda^{\prime}_{ijk} {L}_i {Q}_j {\bar D}_k
+ \lambda_{ijk} {L}_i {L}_j {\bar E}_k
+ \mu_i L_i H,
\label{Rviolation}
\end{equation}
where ${Q}_i, {\bar U}_i, {\bar D}_i, L_i$ and ${\bar E}_i$ denote chiral superfields corresponding to quark doublets, antiquarks, lepton doublets and antileptons, respectively, with $i, j, k$ generation indices. The simultaneous presence of terms of the first and third type in Eq.~(\ref{Rviolation}), namely $\lambda$ and $\lambda^{\prime \prime}$, is severely restricted by lower limits on the proton lifetime, but other combinations are less restricted. The trilinear couplings in Eq.~(\ref{Rviolation}) generate sparticle decays such as ${\tilde q} \to {\bar q} {\bar q}$ or $q \ell$, or ${\tilde \ell} \to \ell \ell$, whereas the bilinear couplings in Eq.~(\ref{Rviolation}) generate Higgs-slepton mixing and thereby also ${\tilde q} \to q \ell$ and ${\tilde \ell} \to \ell \ell$ decays~\cite{Mitsou:2014pua}. For a nominal sparticle mass $\sim 1$~TeV, the lifetime for such decays would exceed a few nanoseconds for $\lambda,  \lambda^{\prime}, \lambda^{\prime \prime} < 10^{-8}$. 

If $R$~parity is broken, the LSP would be unstable, and might be charged and/or coloured. In the former case, it might be detectable directly at the LHC as a massive slowly-moving charged particle. In the latter case, the LSP would bind with light quarks and/or gluons to make colour-singlet states, the so-called \emph{R-hadrons}, and any charged state could again be detectable as a massive slowly-moving charged particle. If $\lambda \ne 0$, the prospective experimental signature would be similar to a stau next-to-lightest sparticle (NLSP) case to be discussed later. On the other hand, if $\lambda^{\prime}$ or $\lambda^{\prime \prime} \ne 0$, the prospective experimental signature would be similar to a stop NLSP case, yielding the possibility of charge-changing interactions while passing through matter. This could yield  a metastable charged particle, created whilst passing through the material surrounding the intersection point,  that would be detected by MoEDAL. 

\subsection{Metastable lepton NLSP in the CMSSM with a neutralino LSP}

However, even if $R$~parity {\it is} exact, the NLSP may be long lived. This would occur, for example, if the LSP is the gravitino, or if the mass difference between the NLSP and the neutralino LSP is small, offering more scenarios for long-lived charged sparticles. In {\it neutralino dark matter} scenarios based on the constrained MSSM (CMSSM), for instance, the most natural candidate for the NLSP is the lighter stau slepton ${\tilde \tau_1}$~\cite{stauNLSP}, which could be long lived if $m_{\tilde \tau_1} - m_{\tilde{\chi}_1^0}$ is small. There are several regions of the CMSSM parameter space that are compatible with the constraints imposed by unsuccessful searches for sparticles at the LHC, as well as the discovered Higgs boson mass. These include a strip in the focus-point region where the relic density of the LSP is brought down into the range allowed by cosmology because of its relatively large Higgsino component, a region where the relic density is controlled by rapid annihilation through direct-channel heavy Higgs resonances, and a strip where the relic LSP density is reduced by coannihilations with near-degenerate staus and other sleptons. It was found in a global analysis that the two latter possibilities are favoured~\cite{MC8}.

In the coannihilation region of the CMSSM, the lighter ${\tilde \tau_1}$ is expected to be the lightest slepton~\cite{stauNLSP}, and the $\tilde\tau_1-\tilde{\chi}_1^0$ mass difference may well be smaller than $m_\tau$: indeed, this is required at large LSP masses. In this case, the dominant stau decays for $m_{\tilde \tau_1} - m_{\tilde{\chi}_1^0} > 160$~MeV are expected to be into three particles: $\tilde{\chi}_1^0 \nu \pi$ or $\tilde{\chi}_1^0 \nu \rho$. If $m_{\tilde \tau_1} - m_{\tilde{\chi}_1^0} < 1.2$~GeV, the ${\tilde \tau_1}$ lifetime is calculated to be so long, in excess of $\sim 100$~ns, that it is likely to escape the detector before decaying, and hence would be detectable as a massive, slowly-moving charged particle~\cite{Sato}. 

\subsection{Metastable sleptons in gravitino LSP scenarios}

\begin{figure}[ht]
\centering\sidecaption
\includegraphics[width=0.5\textwidth]{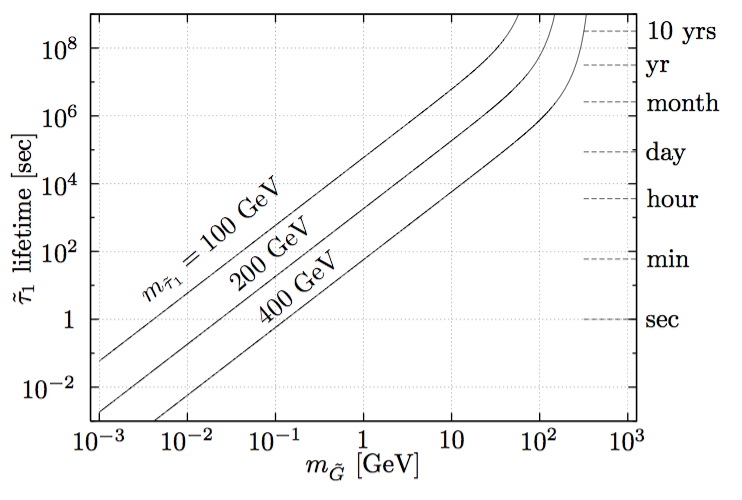}
\caption{The stau lifetime in a gravitino LSP scenario, for different values of the stau mass $m_{\tilde \tau_1}$ and the gravitino mass $m_{\tilde G}$~\cite{CAPTURE3}.}
\label{fig:susy3}
\end{figure}

On the other hand, in {\it gravitino dark matter} scenarios with more general options for the pattern of supersymmetry breaking, other options appear quite naturally, including the lighter selectron or smuon, or a sneutrino~\cite{sleptonNLSP}, or the lighter stop squark ${\tilde t_1}$~\cite{stopNLSP}. If the gravitino ${\tilde G}$ is the LSP, the decay rate of a slepton NLSP is given by 
\begin{equation}
\Gamma ( {\tilde \ell} \to {\tilde G} \ell) = \dfrac{1}{48 \pi {\tilde M}^2} \dfrac{m_{\tilde \ell}^5}{M_{\tilde G}^2}
\left[ 1 - \dfrac{M_{\tilde G}^2}{m_{\tilde \ell}^2} \right]^{4},
\label{telldecay}
\end{equation}
where ${\tilde M}$ is the Planck scale. Since ${\tilde M}$ is much larger than the electroweak scale, the NLSP lifetime is naturally very long, as seen in Fig.~\ref{fig:susy3}~\cite{CAPTURE3}.

\subsection{Long-lived gluinos in split supersymmetry}

The above discussion has been in the context of the CMSSM and similar scenarios where all the supersymmetric partners of Standard Model particles have masses in the TeV range. Another scenario is ``split supersymmetry'', in which the supersymmetric partners of quarks and leptons are very heavy, of a scale $m_s$, whilst the supersymmetric partners of SM bosons are relatively light~\cite{splitSUSY}. In such a case, the gluino could have a mass in the TeV range and hence be accessible to the LHC, but would have a very long lifetime:
\begin{equation}
\tau \approx 8 \left( \dfrac{m_s}{10^9~{\rm GeV}} \right)^4 \left( \dfrac{1~{\rm TeV}}{m_{\tilde{g}}} \right)^5~{\rm s}.
\label{gluinotau}
\end{equation}
Long-lived gluinos would form long-lived gluino R-hadrons including gluino-gluon (gluinoball) combinations, gluino-$q{\bar q}$ (mesino) combinations and gluino-$qqq$ (baryino) combinations. The heavier gluino hadrons would be expected to decay into the lightest species, which would be metastable, with a lifetime given by Eq.~(\ref{gluinotau}), and it is possible that this metastable gluino hadron could be charged.

In the same way as stop hadrons, gluino hadrons may flip charge through conventional strong interactions as they pass through matter, and it is possible that one may pass through most of a conventional LHC tracking detector undetected in a neutral state before converting into a metastable charged state that could be detected by MoEDAL. 

\section{Doubly-charged Higgs bosons}\label{sc:lrsm} 

Doubly-charged particles appear in many theoretical scenarios beyond the SM. For example, doubly-charged scalar states, usually termed doubly-charged Higgs fields, appear in left-right symmetric models~ \cite{Pati1974yy,LRSM,LRSMa} and in see-saw models for neutrino masses with Higgs triplets. A number of models encompasses additional symmetries and extend the SM Higgs sector by introducing doubly-charged Higgs bosons. A representative example of such a model is the L-R Symmetric Model (LRSM)~\cite{Pati1974yy,LRSM,LRSMa}, proposed to amend the fact that the weak-interaction couplings are strictly left handed by extending the gauge group of the SM so as to include a right-handed sector. The simplest realisation is an LRSM~\cite{Pati1974yy, LRSM}  that postulates  a right-handed version of the weak interaction, whose gauge symmetry is spontaneously broken at high mass scale, leading to the parity-violating  SM. This model naturally accommodates recent data on neutrino oscillations and the existence of small neutrino masses. The model generally requires Higgs triplets containing doubly-charged Higgs bosons ($H^{\pm\pm}$)  $\Delta_{R}^{++}$ and $\Delta_{L}^{++}$, which could be light in the minimal supersymmetric left-right model~\cite{LRSUSY}.

Single production of a doubly-charged Higgs boson at the LHC proceeds via vector boson fusion, or through the fusion of a singly-charged Higgs boson with either a $W^\pm$ or another singly-charged Higgs boson. The amplitudes of the  $W_{L} W_{L}$ and $W_{R} W_{R}$ vector boson fusion processes are proportional to $v_{L,R}$, the vacuum expectation values of the neutral members of the scalar triplets of the  LRSM. For the case of $\Delta_{R}^{++}$ production, the vector boson fusion process dominates. Pair production of doubly-charged Higgs bosons is also possible via a Drell-Yan process, with $\gamma$, $Z$ or $Z_{R}$ exchanged in the $s$-channel, but at a high kinematic price since substantial energy is required to produce two heavy particles. In the case of $\Delta_{L}^{++}$, double production may nevertheless be the only possibility if $v_{L}$ is very small or vanishing.

\begin{figure}[ht]
\centering
\sidecaption
\includegraphics[width=0.42\textwidth]{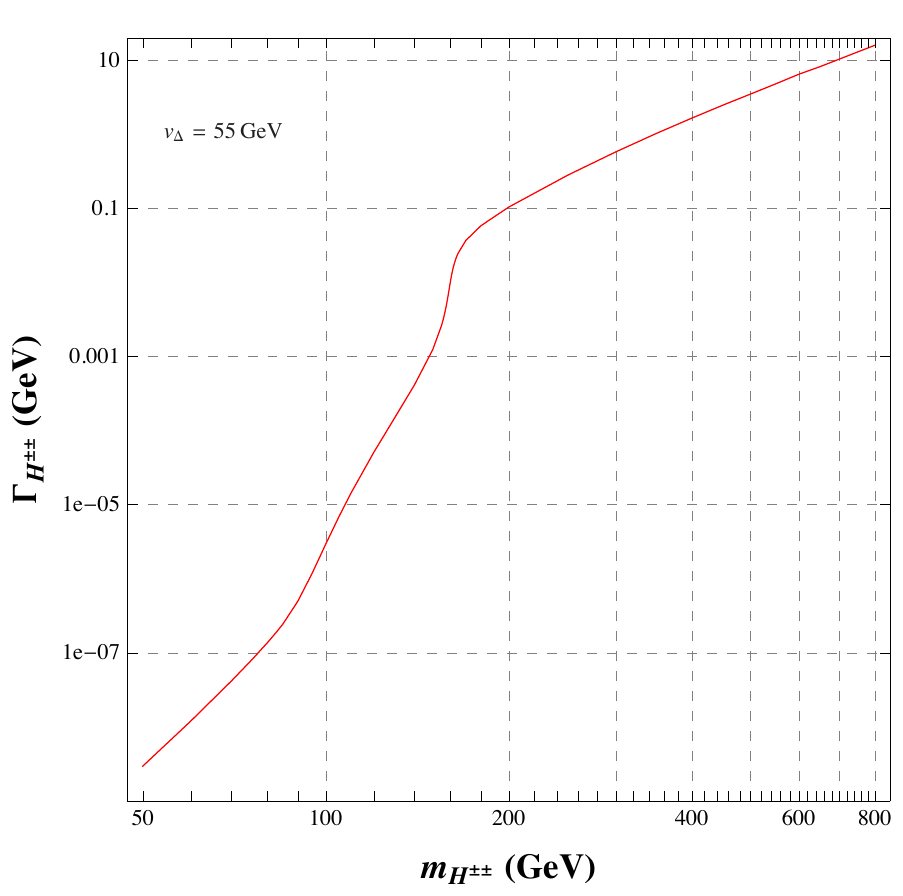}
\caption{Partial decay width of $H^{\pm \pm} \rightarrow W^{\pm} W^{\pm}$ as a function of $m_{H^{\pm\pm}}$ for $v_{\Delta} = 55~{\rm GeV}$~\cite{Chiang:2012dk}.} \label{fig:width}
\end{figure}

The decay of a doubly-charged Higgs boson can proceed via several channels. The dilepton signature leads to the (experimentally clean) final state $q\bar{q} \rightarrow \Delta^{++}_L\Delta^{--}_L \rightarrow 4\ell$. However as long as the triplet vacuum expectation value, $v_\Delta$, is much larger than $10^{-4}~{\rm GeV}$, the doubly-charged Higgs decay predominantly into a pair of same-sign $W$ bosons. For very small Yukawa couplings $H_{\ell\ell} \lesssim  10^{-8}$, the doubly-charged Higgs boson can be quasi-stable. In Fig.~\ref{fig:width}, the partial decay width of the doubly charged Higgs boson into a $W$ boson pair is shown as a function of its mass. For $v_\Delta \gg 10^{-4}~{\rm GeV}$, this partial width is roughly equal to the total width of the doubly charged Higgs boson. In the case of long lifetimes, slowly moving pseudo-stable Higgs bosons could be detected in the MoEDAL NTDs. For example with CR39, one could detect doubly-charged Higgs particles moving with speeds less than around $\beta \simeq 0.4$. 

\section{Extra dimensions}\label{sc:ed}

Over the last decades, models based on compactified extra spatial dimensions (ED) have been proposed in order to explain the large gap between the electroweak (EW) and the Planck scale of $M_{\rm EW}/M_{\rm PL}  \approx 10^{-17}$. The four main scenarios relevant for searches at LHC the Arkani-Hamed-Dimopoulos-Dvali (ADD) model of large extra dimensions~\cite{ADD}, the Randall-Sundrum (RS) model of warped extra dimensions~\cite{Randall}, TeV$^{-1}$-sized extra dimensions~\cite{TEV-1}, and the Universal Extra Dimensions (UED) model~\cite{UED}.

The existence of extra spatial dimensions~\cite{ADD,Randall} and a sufficiently small fundamental scale of gravity open up the possibility that microscopic black holes be produced and detected~\cite{ED1, bhevaporation, fischler1, CHARYBDIS, ED2,ED3}  at the LHC. Once produced, the black holes will undergo an evaporation process categorised in three stages~\cite{bhevaporation, fischler1}: the \emph{balding phase}, the actual \emph{evaporation phase}, and finally   the Planck phase. It is generally assumed that the black hole will decay completely to some last few SM particles. However, another intriguing possibility is that the remaining energy is carried away by a stable remnant.

\subsection{Microscopic black hole remnants}

The prospect of microscopic black hole production at the LHC within the framework of models with large extra dimensions has been studied in Ref.~\cite{ADD}. Black holes produced at the LHC are expected to decay with an average multiplicity of $\sim10-25$ into SM particles,  most of which will be charged, though the details of the multiplicity distribution depend on the number of extra dimensions~\cite{BHMULT}. After the black holes have evaporated off enough energy to reach the remnant mass, some will have accumulated a net electric charge. According to purely statistical considerations, the probability for being left with highly-charged black hole remnants drops fast with the deviation from the average. The largest fraction of the black holes should have charges $\pm1$ or zero, although a smaller but non-negligible fraction would be multiply charged.

The fraction of charged black-hole remnants has been estimated~\cite{BHMULT,Hossenfelder:2005ku} using the {\tt PYTHIA} event generator~\cite{PYTHIA} and the {\tt CHARYBDIS} program~\cite{CHARYBDIS}. It was  assumed that the effective temperature of the black hole drops towards zero for a finite remnant mass, $M_{R}$. The value of $M_{R}$ does not noticeably affect the investigated charge distribution, as it results from the very general statistical distribution of the charge of the emitted particles.

\begin{figure}[htb]
\centering
\sidecaption
\includegraphics[width=0.42\textwidth]{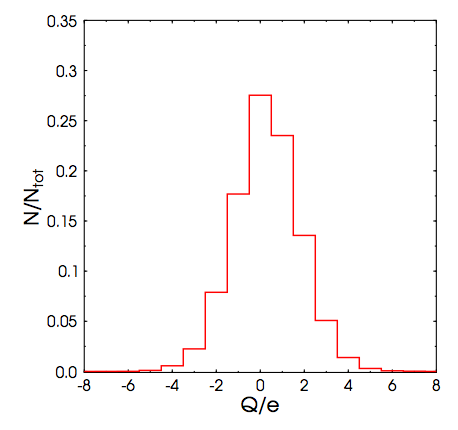}
\caption{The distribution of black-hole remnant charges in proton-proton interactions at $\sqrt{s} = 14~{\rm TeV}$ calculated with the
{\tt PYTHIA} event generator~\cite{PYTHIA} and the {\tt CHARYBDIS} program~\cite{CHARYBDIS}. From Ref.~\cite{Hossenfelder:2005ku}.} \label{Fig:Qofremnants}
\end{figure}

Thus, independent of the underlying quantum-gravitational  assumption leading to the remnant formation, it was found that about 30\% of the remnants are neutral, whereas $\sim$ 40\% would be singly-charged black holes, and the  remaining $\sim$30\% of remnants would be multiply-charged.  The distribution of the  remnant charges obtained  is shown in Fig.~\ref{Fig:Qofremnants}. The black hole remnants  considered here are heavy, with masses of a TeV or more. A significant  fraction of the black-hole remnants produced would have a  Z/$\beta$ of greater than five, high enough to register in the CR39 NTDs forming the LT-NTD  sub-detector of MoEDAL.

\subsection{D-matter}\label{sc:dm}

Some versions of string theory include higher-dimensional ``domain-wall''-like membrane \emph{(brane)} structures in space-time. Fundamental open strings, which represent particle excitations of the vacuum, have their ends attached to membranes embedded in higher-spatial dimensional \emph{bulk} spaces. On the other hand, only gravity and closed string  modes are free to propagate in the bulk space between branes. These brane structures are called \emph{D-branes} since the attachment of the ends of the open strings is described by Dirichlet world-sheet boundary conditions.

Nevertheless, the notion of higher-dimensional space-times with domain-world structures leads to cases where the bulk is ``punctured'' by lower-dimensional D-brane defects, which are either point-like or have their longitudinal dimensions compactified~\cite{westmuckett}. From a low-energy observer's perspective, living on a brane Universe with three spatial longitudinal, uncompactified dimensions, such structures would effectively appear to be point-like \emph{D-particles}. The latter have  dynamical degrees of freedom, thus they can be treated as quantum  excitations above the vacuum~\cite{westmuckett,shiu} collectively referred to as {\it D-matter}. D-matter states are non-perturbative stringy objects with masses of order $m_D \sim M_s/g_s$, where $g_s $ is the string coupling, typically of order one so that the observed gauge and gravitational couplings is reproduced. Hence, the D-matter states could be light enough to be phenomenologically relevant at the LHC.

Depending on their type, D-branes could carry integral or torsion (discrete) charges. The lightest D-particle (LDP) is stable being the lightest state carrying this particular charge. Therefore, just as in the case of the SUSY LSP, the LDPs are possible candidates for cold dark matter~\cite{shiu}. D-particles, like all other D-branes, are solitonic non-perturbative objects in the string/brane theory. As discussed in the relevant literature~\cite{shiu}, there are similarities and differences between D-particles and magnetic monopoles with non-trivial cosmological implications~\cite{Witten2002wb,westmuckett,Mavromatos:2010jt,mitsou}. 

An important difference between the D-matter states and other non-perturbative objects, such as magnetic monopoles, is that they could have {\it perturbative } couplings, with no magnetic charge in general. Nonetheless, in the context of brane-inspired gauge theories, brane states with magnetic charges can be constructed, which would manifest themselves in MoEDAL in a manner similar to  magnetic monopoles. 
 
Non-magnetically-charged  D-matter, on the other hand, could be produced at colliders and also produce interesting signals of direct relevance to the MoEDAL experiment. For instance, excited states of D-matter (${\rm D}^\star$) can be electrically-charged, provided one end of the open string is attached to the D3 brane Universe. Such charged states are supermassive when compared to the SM states. For typical string couplings of phenomenological relevance, the first few massive levels may be accessible to the LHC.  Depending on the details of the microscopic model considered, and the way the SM is embedded, such massive charged states can be relatively long-lived, thereby playing a role  analogous to the possible long-lived charged particles in low-energy supersymmetric models, and could likewise be detectable with MoEDAL. 

\begin{figure}[ht]
\centering
\subfigure[Signal]{ 
\includegraphics[width=0.33\textwidth]{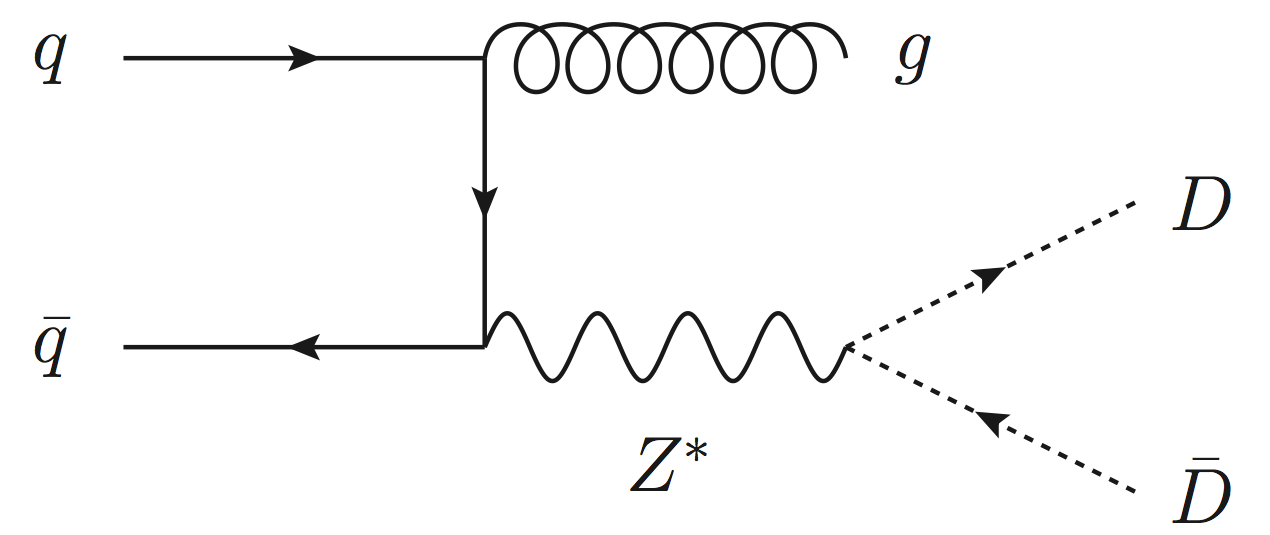}
\label{sfig:signal}
}
\quad
\subfigure[DM production]{
\includegraphics[width=0.33\textwidth]{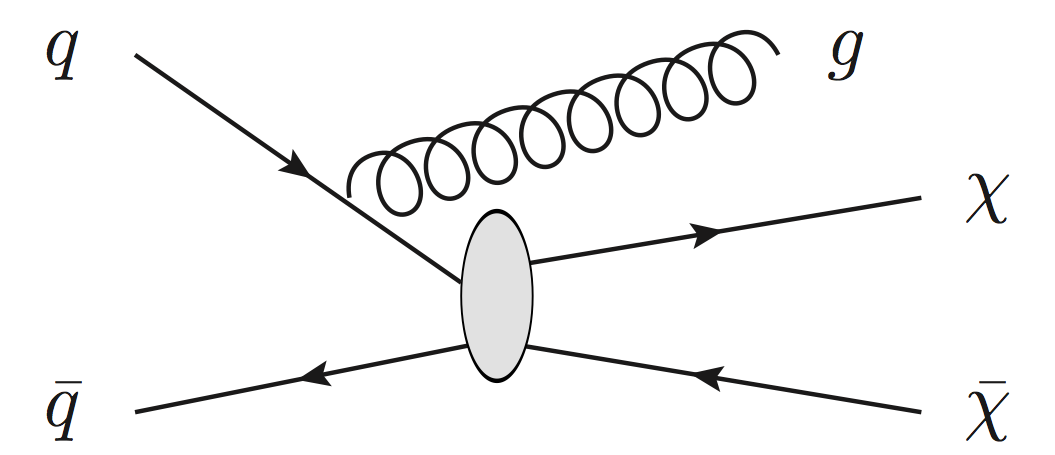}
\label{sfig:DM-prod}
}
~~~~
\subfigure[$Z\to\nu\nu$ background]{
\includegraphics[width=0.33\textwidth]{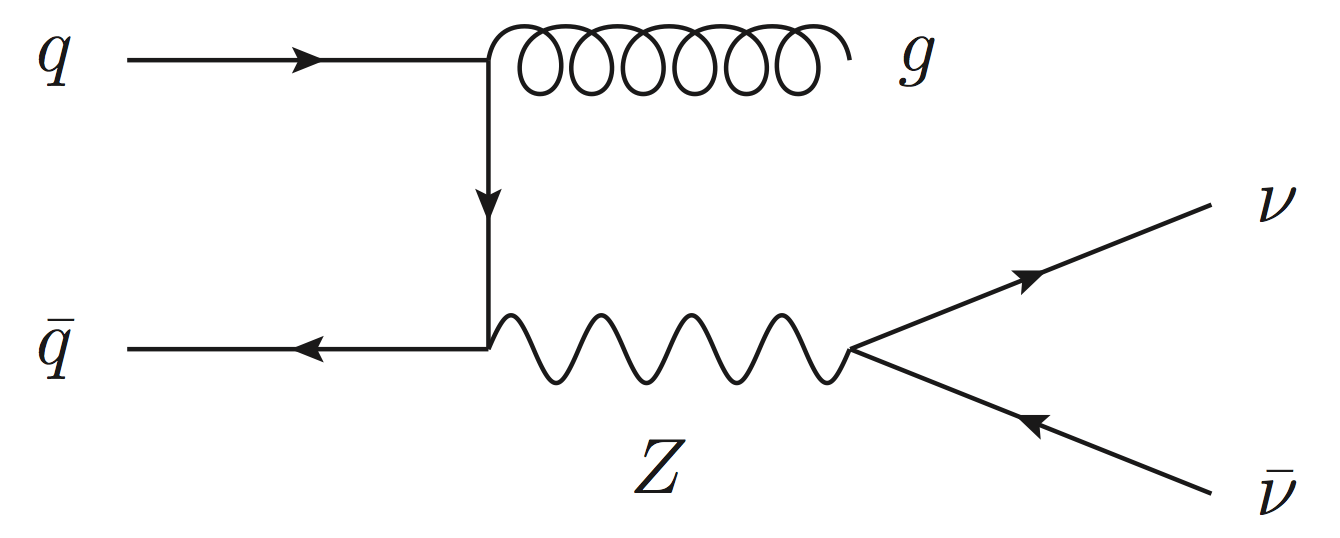}
\label{sfig:bkg1}
}
\quad
\subfigure[$W\to\ell_{\rm inv}\nu$ background]{
\includegraphics[width=0.33\textwidth]{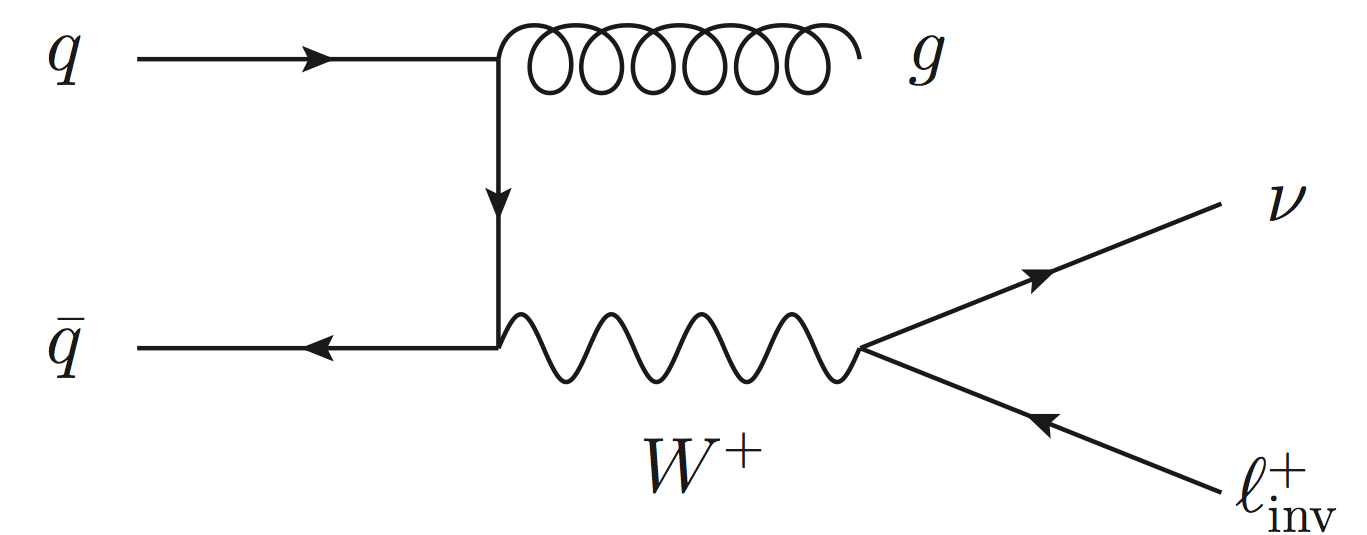}
\label{sfig:bkg2}
}
\caption{(a) An example of parton-level diagrams for production of D-particles by $q\bar{q}$ collisions in a generic D-matter low-energy model~\cite{mitsou}. (b) Production of conventional dark-matter particle-antiparticle ($\chi\bar{\chi}$) in effective field theories, assuming that dark matter, co-existing with D-matter, couples to quarks via higher-dimensional contact interactions~\cite{kolbmaverick,lhcmaverick}.  (c) \& (d) The dominant SM background processes.}
\label{fig:dproduction1}
\end{figure}

D-matter/antimatter pairs can be produced~\cite{Mavromatos:2010jt,mitsou} by the decay of intermediate off-shell $Z$-bosons. The D-matter pairs produced in a hadron collider will traverse the detector and exit undetected, as they are only weakly interacting, giving rise to large transverse missing energy, $E{\rm _T^{miss}}$. Hence mono-$X$ analyses, targeting DM-pair production plus an initial-state-radiation jet, photon or gauge boson, such as the one shown in Fig.~\ref{sfig:DM-prod}, would be of highly relevant investigative tool. 
 
The dominant SM background in such searches involves the decay of a $Z$ to a neutrino pair and of $W^{\pm}$ to an ``invisible'' ($\ell_{\rm inv}$) lepton and a neutrino, as depicted in  Figs.~\ref{sfig:bkg1} and~\ref{sfig:bkg2}, respectively. The null results found using, e.g., the ATLAS detector in searches for missing energy + jets, impose a bound on the dark matter mass $m_\chi$ and the D-particle coupling $g_{37}$. With the LHC data at $\sqrt{s}=8$~TeV and 20~fb$^{-1}$ integrated luminosity, the current lower bounds on the DM mass set by ATLAS using a mono-$W/Z$ analysis is ${\cal O}(1$~TeV)~\cite{atlas-dm2014}. On the other hand, as mentioned previously, excited states of such a LDP, of mass $M_{D^\star}$, involving stretched  strings between the D-particle and the brane world, can be charged and thus highly-ionising. Thus, they  are of relevance to the MoEDAL detector searches, provided the string scale is sufficiently low. 

\section{Summary and outlook}\label{sc:summary}

MoEDAL is going to extend considerably the LHC reach in the search for (meta)stable highly ionising particles. The latter are predicted in a variety of theoretical models: magnetic monopoles, SUSY stable spartners, quirks, strangelets, Q-balls, etc~\cite{Acharya:2014nyr}. The MoEDAL design is optimised to probe precisely such long lived states, unlike the other LHC experiments~\cite{DeRoeck:2011aa}. Furthermore it combines different detector technologies: plastic nuclear track detectors, trapping volumes and pixel sensors~\cite{moedal-tdr}. The first physics results, obtained with LHC Run~I data, are going to be published soon~\cite{Joergensen:2012gy}. The MoEDAL experiment is currently getting ready for the LHC Run~II due to start in 2015.

\section*{Acknowledgements}

The author is grateful to the ICNFP2014 organisers for the kind invitation to present this talk in the conference. She acknowledges support by the Spanish Ministry of Economy and Competitiveness (MINECO) under the projects FPA2012-39055-C02-01, by the Generalitat Valenciana through the project PROMETEO~II/2013-017 and by the Spanish National Research Council (CSIC) under the JAE-Doc program co-funded by the European Social Fund (ESF). 



\begin{thebibliography}{99}

\bibitem{moedal-web} For general information on the MoEDAL experiment, see: \url{http://moedal.web.cern.ch/}

\bibitem{moedal-tdr} MoEDAL Collaboration, Technical Design Report of the MoEDAL Experiment {\it CERN Preprint} CERN-LHC-2009-006, MoEDAL-TDR-1.1 (2009), and references therein.

\bibitem{jim}   J.~L.~Pinfold,
  ``The MoEDAL Experiment -- a New Light on LHC Physics ,'' these proceedings.
  
\bibitem{LHC}  L.~Evans and P.~Bryant,
  JINST {\bf 3}, S08001 (2008).

\bibitem{DeRoeck:2011aa}  A.~De Roeck, A.~Katre, P.~Mermod, D.~Milstead and T.~Sloan,
  Eur.\ Phys.\ J.\ C {\bf 72}, 1985 (2012). 
  
\bibitem{Fairbairn07} M.~Fairbairn, A.~C.~Kraan, D.~A.~Milstead, T.~Sjostrand, P.~Z.~Skands and T.~Sloan,
  Phys.\ Rept.\  {\bf 438}, 1 (2007);  
S.~Burdin, M.~Fairbairn, P.~Mermod, D.~Milstead, J.~Pinfold, T.~Sloan and W.~Taylor,
  ``Non-collider searches for stable massive particles,''
  arXiv:1410.1374 [hep-ph] (2014).
  
\bibitem{Acharya:2014nyr}    B.~Acharya {\it et al.}  [MoEDAL Collaboration],
  Int.\ J.\ Mod.\ Phys.\ A {\bf 29}, 1430050 (2014), and references therein. 

\bibitem{LHCb-detector}   A.~A.~Alves, Jr. {\it et al.}  [LHCb Collaboration],
  JINST {\bf 3}, S08005 (2008).
  
\bibitem{Joergensen:2012gy}    M.~D.~Joergensen, A.~De Roeck, H.-P.~Hachler, A.~Hirt, A.~Katre, P.~Mermod, D.~Milstead and T.~Sloan,
  ``Searching for magnetic monopoles trapped in accelerator material at the Large Hadron Collider,''
  arXiv:1206.6793 [physics.ins-det] (2012).
  
\bibitem{DeRoeck:2012wua}   A.~De~Roeck, H.~P.~Hachler, A.~M.~Hirt, M.-D.~Joergensen, A.~Katre, P.~Mermod, D.~Milstead and T.~Sloan,
  Eur.\ Phys.\ J.\ C {\bf 72}, 2212 (2012).
  
\bibitem{Dirac1931kp} P.~A.~M.~Dirac,
  Proc.\ Roy.\ Soc.\ Lond.\ A {\bf 133}, 60 (1931). 

\bibitem{Diracs_idea} P.~A.~M.~Dirac, Phys.\ Rev.\ {\bf 74}, 817 (1948).

\bibitem{tHooft-Polyakov}
G.~'t~Hooft, Nucl.\ Phys.\ B {\bf 79}, 276 (1974); 
A.~M.~Polyakov, JETP Lett.\  {\bf 20}, 194 (1974).

\bibitem{mm}
B.~Julia and A.~Zee, 
  Phys.\ Rev.\ D {\bf 11}, 2227 (1975); 
Y.~Nambu, Nucl.\ Phys.\ B {\bf 130}, 505 (1977); 
E.~Witten,
  Phys.\ Lett.\ B {\bf 86}, 283 (1979); 
G.~Lazarides, M.~Magg  and  Q.~Shafi,
  Phys.\ Lett.\ B {\bf 97}, 87 (1980);
R.~D.~Sorkin,
  Phys.\ Rev.\ Lett.\ {\bf 51}, 87 (1983); 
D.~J.~Gross  and  M.~J.~Perry, 
  Nucl.\ Phys.\ B {\bf 226}, 29 (1983); 
J.~S.~Schwinger,
  Science {\bf 165}, 757 (1969); 
J.~Preskill,
  Ann.\ Rev.\ Nucl.\ Part.\ Sci.\ {\ bf 34}, 461 (1984); 
A.~Achucarro and T.~Vachaspati, Phys.\ Rep.\ {\bf 327}, 347  (2000); 
T.~W.~Kephart, C.-A.~Lee  and Q.~Shafi,
 JHEP {\bf 0701}, 088 (2007); 
Y.~M.~Cho, K.~Kim and J.~H.~Yoon, 
 ``Mass of the Electroweak Monopole,'' arXiv:1212.3885 [hep-ph] (2012); 
D.~G.~Pak, P.~M.~Zhang and L.~P.~Zou,
  ``Screened monopoles in Weinberg-Salam model,''
  arXiv:1311.7567 [hep-th] (2013); 
A.~Rajantie, 
  JHEP {\bf 0601}, 088 (2006). 

\bibitem{Cho1996qd}  Y.~M.~Cho and D.~Maison,
  Phys.\ Lett.\ B {\bf 391}, 360 (1997);  
  W.~S.~Bae and Y.~M.~Cho, JKPS {\bf 46}, 791 (2005).

\bibitem{Vento2013jua}  V.~Vento and V.~S.~Mantovani,
  ``On the magnetic monopole mass,''
  arXiv:1306.4213 [hep-ph] (2013).

\bibitem{khlopov} Y.~B.~Zeldovich and M.~Y.~Khlopov, Phys.\ Lett.\ B  {\bf 79},  239 (1978).

\bibitem{Monopolium} C.~T.~Hill, Nucl.\ Phys.\ B {\bf 224}, 469 (1983). 

\bibitem{Monopolium1} V.~K.~Dubrovich, Grav.\ Cosmol.\ Suppl.\ {\bf 8 N1},  122  (2002).

\bibitem{Rajantie:2012xh}   A.~Rajantie,
  Contemp.\ Phys.\ {\bf  53}, 195 (2012).  

\bibitem{Dirac:1931kp}  P.~A.~M.~Dirac,
  Proc.\ Roy.\ Soc.\ Lond.\  A {\bf 133} (1931) 60.

\bibitem{Epele0}
L.~N.~Epele, H.~Fanchiotti, C.~A.~Garcia Canal and V.~Vento, Eur.\ Phys.\ J.\ C {\bf 56},  87  (2008); 
Eur.\ Phys.\ J.\ C {\bf 62}, 587 (2009).     

\bibitem{Epele1} L.~N.~Epele, H.~Fanchiotti, C.~A.~Garcia Canal, V.~A.~Mitsou and V.~Vento, Eur.\ Phys.\ J.\ Plus {\bf 127}, 60 (2012).


\bibitem{susy} H.~P.~Nilles, 
  Phys.\ Rept.\ {\bf 110} (1984) 1.

\bibitem{EHNOS} H.~Goldberg,
   Phys.\ Rev.\ Lett.\ {\bf 50}, 1419  (1983); 
J.~Ellis, J.~Hagelin, D.~Nanopoulos, K.~Olive and M.~Srednicki,
   Nucl.\ Phys.\ B {\bf 238}, 453 (1984).


\bibitem{RV}  R.~Barbieri {\it et al.},
  Phys.\ Rept.\  {\bf 420}, 1 (2005).

\bibitem{Mitsou:2014pua} 
  V.~A.~Mitsou,
  EPJ Web Conf.\  {\bf 71}, 00092 (2014), 
    
  
\bibitem{stauNLSP}  J.~R.~Ellis, K.~A.~Olive, Y.~Santoso and V.~C.~Spanos,
Phys.\ Lett.\ B {\bf 565}, 176 (2003). 

\bibitem{MC8} O.~Buchmueller {\it et al.},
  Eur.\ Phys.\ J.\ C {\bf 72}, 2243  (2012);  
 {\it ibid.\ }  {\bf 74}, 2922 (2014). 
    
\bibitem{Sato}  T.~Jittoh, J.~Sato, T.~Shimomura and M.~Yamanaka,
  Phys.\ Rev.\ D {\bf 73}, 055009 (2006). 
    
\bibitem{sleptonNLSP} J.~R.~Ellis, K.~A.~Olive and Y.~Santoso,
  JHEP {\bf 0810}, 005  (2008). 

\bibitem{stopNLSP}  J.~L.~Diaz-Cruz, J.~R.~Ellis, K.~A.~Olive and Y.~Santoso,
  JHEP {\bf 0705}, 003  (2007). 
    
\bibitem{CAPTURE3} K.~Hamaguchi, M.~M.~Nojiri and A.~de Roeck,
  JHEP {\bf 0703}, 046 (2007). 

  
    
  
\bibitem{splitSUSY}  N.~Arkani-Hamed and S.~Dimopoulos, JHEP {\bf 0506}, 073 (2005);
G.~F.~Giudice and A.~Romanino,  
Nucl.\ Phys.\ B {\bf 699}, 65 (2004) [Erratum-ibid.\ B {\bf 706}, 65 (2005)].

\bibitem{Pati1974yy} J.~C.~Pati and A.~Salam, 
 Phys.\ Rev.\ D {\bf 10}, 275 (1974) [Erratum-ibid D 11, 703 (1975)].
 
\bibitem{LRSM} R.~N.~Mohapatra and J.~C.~Pati, Phys.\ Rev.\ D {\bf 11}, 566 (1975).

\bibitem{LRSMa} G.~Senjanovic and R.~N.~Mohapatra, Phys.\ Rev.\ D {\bf 12}, 1502 (1975); 
R.~N.~Mohapatra and G.~Senjanovic, Phys.\ Rev.\ D {\bf 23}, 165 (1981). 


\bibitem{LRSUSY} C.~S.~Aulakh, A.~Melfo, and G.~Senjanovic, Phys.\ Rev.\ D {\bf 57}, 4174 (1998); 
Z.~Chacko and R.~N.~Mohapatra, Phys.\ Rev.\ D {\bf 58}, 015003 (1998); 
C.~S.~Aulakh, K.~Benakli and G.~Senjanovic, Phys.\ Rev.\ Lett.\ {\bf 79}, 2188 (1997). 

\bibitem{Chiang:2012dk} C.~W.~Chiang, T.~Nomura and K.~Tsumura,
  Phys.\ Rev.\ D {\bf 85}, 095023 (2012). 

\bibitem{ADD} N.~Arkani-Hamed, S.~Dimopoulos and G.~R.~Dvali, Phys.\ Lett.\ B {\bf 429}, 263 (1998); 
Phys.\ Rev.\ D {\bf 59}, 086004 (1999); \\
I.~Antoniadis,  N.~Arkani-Hamed, S.~Dimopoulos and G.~R.~Dvali, Phys.\ Lett.\ B {\bf 436}, 257 (1998).

\bibitem{Randall} L.~Randall and R.~Sundrum, Phys.\ Rev.\ Lett.\ {\bf 83}, 4690 (1999).

\bibitem{TEV-1} I.~Antoniadis,  Phys.\ Lett.\ B {\bf 377}, 246 (1990); 
I.~Antoniadis and K.~Benakli,  Phys.\ Lett.\ B {\bf 326}, 69 (1994); 
I.~Antoniadis, K.~Benakli and M.~Quiros, Phys.\ Lett.\ B {\bf 331}, 313 (1994). 

\bibitem{UED} T.~Appelquist, H.-Ch.~Cheng and B.~A.~Dobrescu, Phys.\ Rev.\ D {\bf 64}, 035002 (2001). 

\bibitem{bhevaporation} S.~B.~Giddings and S.~Thomas, Phys.\ Rev.\ D {\bf 65}, 056010 (2002).

\bibitem{fischler1} W.~Fischler, ``A Model for high-energy scattering in quantum gravity,'' arXiv:hep-th/9906038 (1999).

\bibitem{ED1} P.~C.~Argyres, S.~Dimopoulos and J.~March-Russell, Phys.\ Lett.\ B {\bf 441}, 96 (1998); 
  S.~Dimopoulos and G.~L.~Landsberg, Phys.\ Rev.\ Lett.\  {\bf 87}, 161602 (2001); 
  G.~L.~Alberghi, R.~Casadio and A.~Tronconi, J.\ Phys.\ G {\bf 34}, 767 (2007); 
  M.~Cavaglia, R.~Godang, L.~Cremaldi and D.~Summers, Comput.\ Phys.\ Commun.\ {\bf 177}, 506 (2007); 
  D.~C.~Dai, G.~Starkman, D.~Stojkovic, C.~Issever, E.~Rizvi and J.~Tseng, Phys.\ Rev.\ D {\bf 77}, 076007 (2008).

\bibitem{ED2} R.~Casadio and B.~Harms, Int.\ J.\ Mod.\ Phys.\ A {\bf 17}, 4635 (2002).

\bibitem{ED3} M.~Cavaglia, Int.\ J.\ Mod.\ Phys.\ A {\bf 18}, 1843 (2003); 
  P.~Kanti, Int.\ J.\ Mod.\ Phys.\ A {\bf 19}, 4899 (2004).

\bibitem{CHARYBDIS} C.~M.~Harris, P.~Richardson and B.~R.~Webber, JHEP {\bf 0308}, 033 (2003).  

\bibitem{BHMULT} B.~Koch, M.~Bleicher and H.~Stoecker, J.\ Phys.\ G \textbf{34}, S535 (2007).

\bibitem{Hossenfelder:2005ku}  S.~Hossenfelder, B.~Koch and M.~Bleicher,
  ``Trapping black hole remnants,'' hep-ph/0507140 (2005).
  
\bibitem{PYTHIA} T.~Sjostrand, S.~Mrenna and P.~Z.~Skands,
  JHEP {\bf 0605} (2006) 026. 
  
\bibitem{westmuckett} J.~R.~Ellis, N.~E.~Mavromatos and D.~V.~Nanopoulos,
  Gen.\ Rel.\ Grav.\  {\bf 32}, 943 (2000); 
Phys.\ Lett.\ B {\bf 665}, 412 (2008); 
J.~R.~Ellis, N.~E.~Mavromatos and M.~Westmuckett, 
Phys.\ Rev.\ D \textbf{70}, 044036 (2004);  
{\it ibid.\ } \textbf{71}, 106006 (2005).
   
\bibitem{shiu}  G.~Shiu and L.~-T.~Wang,
  Phys.\ Rev.\ D {\bf 69}, 126007 (2004).

\bibitem{Mavromatos:2010jt} 
  N.~E.~Mavromatos, S.~Sarkar and A.~Vergou,
  Phys.\ Lett.\ B {\bf 696}, 300 (2011). 
  
\bibitem{mitsou}  N.~E.~Mavromatos, V.~A.~Mitsou, S.~Sarkar and A.~Vergou,
  Eur.\ Phys.\ J.\ C {\bf 72}, 1956 (2012).
 
\bibitem{Witten2002wb}  See, e.g., E.~Witten,
  ``Comments on string theory,''
  arXiv:hep-th/0212247 (2002), and references therein.  

\bibitem{kolbmaverick}  M.~Beltran, D.~Hooper, E.~W.~Kolb, Z.~A.~C.~Krusberg and T.~M.~P.~Tait,
  JHEP {\bf 1009}, 037 (2010). 

\bibitem{lhcmaverick}  A.~Rajaraman, W.~Shepherd, T.~M.~P.~Tait and A.~M.~Wijangco,
  Phys.\ Rev.\ D {\bf 84}, 095013 (2011). 
  
\bibitem{atlas-dm2014}  G.~Aad {\it et al.} [ATLAS Collaboration],
  Phys.\ Rev.\ Lett.\  {\bf 112}, 041802 (2014). 
  
  
\end{thebibliography}
\end{document}